\documentstyle[eqsecnum,aps,epsf]{revtex}

\begin{document}


\def\B{{\cal B}}
\def\F{{\cal F}}
\def\H{{\cal H}}
\def\I{{\cal I}}
\def\P{{\cal P}}
\def\T{{\cal T}}
\def\U{{\cal U}}
\newcommand{\bm}[1]{
\mbox{\boldmath$ #1 $}}

\newtheorem{theorem}{Theorem}
\newtheorem{lemma}[theorem]{Lemma}
\newtheorem{definition}[theorem]{Definition}
\newenvironment{proof}{\par\noindent {\bf Proof}\par}{\par\medskip}
\def\QED{$\Box$}


{\baselineskip=0pt
\leftline{\large\sl\vbox to0pt{\hbox{Yukawa Institute Kyoto}\vss}}
\rightline{\large\rm\vbox to0pt{\hbox{YITP-96-38}
				\hbox{\today}\vss}}
}
\vskip15mm

\begin{center}
{\Large New Proof of the Generalized Second Law}

\vskip5mm
{\large Shinji Mukohyama}
\vskip3mm
{Yukawa Institute for Theoretical Physics, Kyoto University \\ 
Kyoto 606-01, Japan}
\end{center}


\begin{abstract}
The generalized second law of black hole thermodynamics was proved by
Frolov and Page for a quasi-stationary eternal black hole. However,
realistic black holes arise from a gravitational collapse, and in this
case their proof does not hold. In this paper we prove the generalized
second law for a quasi-stationary black hole which arises from a
gravitational collapse.
\end{abstract}

\vskip1cm



\section{Introduction} 
The generalized second law of black hole thermodynamics insists that
an entropy of a black hole plus a thermodynamic entropy of fields
outside the horizon does not decrease \cite{GSL}, where the black hole
entropy is defined as a quarter of the area of the horizon. Namely it
says that an entropy of the whole system does not decrease. It
interests us in a quite physical sense since it links a world inside a 
black hole and our thermodynamic world. In particular it gives a
physical meaning to black hole entropy indirectly since it concerns
the sum of black hole entropy and ordinary thermodynamic entropy and
physical meaning of the later is well-known by statistical mechanics.

Frolov and Page \cite{Frolov&Page} proved the generalized second law
for a quasi-stationary eternal black hole by assuming that a state of
matter fields on the past horizon is thermal one and that a set of
radiation modes on the past horizon and one on the past null infinity
are quantum mechanically uncorrelated. The assumption is reasonable
for the eternal case since a black hole emit a thermal radiation (the
Hawking radiation). When we attempt to apply their proof to a
non-eternal black hole which arises from a gravitational collapse, we
might expect that things would go well by simply replacing the past
horizon with a null surface at a moment of a formation of a
horizon ($v=v_0$ surface in {\it Figure}
\ref{fig:background}). However, the expectation is disappointing since
the assumption becomes ill in this case. The reason is that on a
collapsing background the thermal radiation is observed not at the 
moment of the horizon formation but at the future null infinity and
that any modes on the future null infinity have correlation with modes
on the past null infinity located after the horizon formation. The
correlation can be seen in the equation (\ref{eqn:multi}) of this
paper explicitly. Thus, their proof does not hold for the case in
which a black hole arises from a gravitational collapse. Since
astrophysically a black hole is thought to arise from a gravitational
collapse, we want to prove the generalized second law in this case.

In this paper we prove the generalized second law for a
quasi-stationary black hole which arises from a gravitational
collapse. For this purpose we concentrate on an inequality between 
functionals of a density matrix since the generalized second law can
be rewritten as an inequality between functionals of a density matrix
of matter fields as
shown in the Sec. \ref{sec:GSL}. We seek a method to prove that a
special functional of a density matrix cannot decrease under a
physical evolution. (It is a generalization of a result by Sorkin
\cite{Sorkin}.) To apply it to the system with a black hole and derive
the generalized second law as its consequence we need to establish a
property of physical evolution of matter fields around the black
hole. Thus, for concreteness, we investigate a real massless scalar
field semiclassically in a curved background which describes
gravitational collapse and calculate 
conditional probabilities which, as a whole, have almost all
information about behaviors of the scalar field after the formation of
the horizon. (The probability we seek is a generalization of one 
calculated by Panangaden and Wald \cite{Panangaden&Wald}.) Using the
result of the calculation, it is shown that a thermal density matrix
of the scalar field at the past null infinity evolves to a thermal
density matrix with the same temperature and the same chemical
potential at the future null infinity, provided that the initial
temperature and chemical potential are special values specified by the
background geometry. Finally we prove the generalized second law by
using these results. 

The rest of the paper is organized as follows. In
Sec. \ref{sec:scalar} we consider a real massless scalar field in a
background of a gravitational collapse to show a thermodynamic 
property of it. A thermal state with special values of temperature
and chemical potential evolves to a thermal state with the same 
temperature and the same chemical potential. These special values are
determined by the background geometry. In Sec. \ref{sec:GSL},
first the generalized second law is rewritten as an inequality which
states that there is a non-decreasing functional of a density matrix
of matter fields. After that, we give a theorem which shows an
inequality between functionals of density matrices. Finally we apply
it to the scalar field investigated in Sec. \ref{sec:scalar} to prove
the generalized second law for the quasi-stationary background. In
Sec. \ref{sec:summary} we summarize this paper.


\section{massless scalar field in black hole background}
	\label{sec:scalar}

In this section we consider a real massless scalar field in a curved
background which describes a formation of a quasi-stationary black
hole. Let us denote a past null infinity by $\I^-$,a future null
infinity by $\I^+$ and a future event horizon by $H^+$. Introduce
usual null coordinate $u$, $v$ and suppose that the formation of the
event horizon $H^+$ is at $v=v_0$ (see {\it Figure}
\ref{fig:background}). At $\I^-$ and $\I^+$, by virtue of the
asymptotic flatness, there is a natural definition of Hilbert spaces
$\H_{\I^-}$ and $\H_{\I^+}$ of mode functions with positive
frequencies \cite{Wald}. Hilbert spaces $\F(\H_{\I^\pm})$ of all 
asymptotic states are defined as follows with a suitable completion
(symmetric Fock spaces). 
\[
 \F(\H_{\I^\pm}) \equiv 
	\bm{C} \oplus \H_{\I^\pm} \oplus 
	\left(\H_{\I^\pm}\otimes\H_{\I^\pm}\right)_{sym} 
	\oplus \cdots,	
\]
where $(\cdots)_{sym}$ denotes the symmetrization
($(\xi\otimes\eta )_{sym}=
	\frac{1}{2}(\xi\otimes\eta +\eta\otimes\xi )$, etc.). 
Physically, $\bm{C}$ denotes the vacuum state, $\H_{\I^\pm}$ one
particle states, $\left(\H_{\I^\pm}\otimes\H_{\I^\pm}\right)_{sym}$
two particle states, etc.. We suppose that all our observables are 
operators on $\F(\H_{\I^\pm})$ since we observe a radiation of the
scalar field radiated by the black hole at place far away from it. In
this sense $\F(\H_{\I^+})$ are quite physical. Next let us consider
how to set an initial state of the scalar field. We want to see a
response of the scalar field on the quasi-stationary black hole
background which arises from a gravitational collapse of other
materials (a dust, a fluid, etc.). Hence the initial state at $\I^-$
must be such a state that it includes no excitations of modes located
before the formation of the horizon (no excitation at $v<v_0$). A
space of all such states is a subspace of $\F(\H_{\I^-})$, and we
denote it by $\F_{\I^-(v>v_0)}$. We like to derive a thermal property
of a scattering process of the scalar field by the quasi-stationary
black hole. Hence we consider density matrices on $\F_{\I^-(v>v_0)}$
and $\F(\H_{\I^+})$. Denote a space of all density matrices on
$\F_{\I^-(v>v_0)}$ by $\P$ and a space of all density matrices on
$\F(\H_{\I^+})$ by $\tilde{\P}$. 

Let us discuss an evolution of a state at $\I^-$ to future. Since
$\I^{+}$ is not a Cauchy surface because of the existence 
of $H^+$, $\F(\H_{\I^-})$ is mapped not to $\F(\H_{\I^+})$ but to
$\F(\H_{\I^+})\otimes\F(\H_{H^+})$ by a unitary evolution, where
$\H_{H^+}$ is a Hilbert space of mode functions on the horizon with a
positive frequency and $\F(\H_{H^+})$ is a Hilbert
space of all states on $H^+$ defined as a symmetric Fock
space (see the definition of $\F(\H_{\I^\pm})$). Although there is no
natural principle to determine the positivity of the frequency
(equivalently, there is no natural definition of the particle concept), 
the detailed definition of $\H_{H^+}$ does not affect the result since
we shall trace out the degrees of freedom of $\F(\H_{H^+})$ (see
(\ref{eqn:def-T})).  To describe an evolution of a quantum state of
the scalar field from $\F(\H_{\I^-})$ to $\F(\H_{\I^+})\otimes\F(\H_{H^+})$
a S-matrix is introduced \cite{Wald}. For a given initial state
$|\psi\rangle$ in $\F(\H_{\I^-})$, the corresponding final state in
$\F(\H_{\I^+})\otimes\F(\H_{H^+})$ is $S|\psi\rangle$. Then the
corresponding evolution from $\F_{\I^-(v>v_0)}$ to $\F(\H_{\I^+})$ is
obtained by restricting $S$ to $\F_{\I^-(v>v_0)}$, and we denote it by
$S$, too. In this section we show a thermal property of the scalar
field in the background by using the S-matrix elements given by Wald
\cite{Wald}.

\subsection{Definition of $T$}
      \label{subsec:def-T} 
Let the initial state of the scalar field be $|\phi\rangle$
($\in\F(\H_{\I^-})$), and observe the corresponding final state at
$\I^+$ (see the argument after the definition of
$\F(\H_{\I^\pm})$). Formally the observation corresponds to a
calculation of a matrix element 
$\langle\phi |S^{\dagger}OS|\phi\rangle$, where $S$ is the
S-matrix which describes the evolution of the scalar field from
$\F_{\I^-(v>v_0)}$ to $\F(\H_{\I^+})\otimes\F(\H_{H^+})$ and $O$ is a
self-adjoint operator on $\F(\H_{\I^+})$ corresponding to a quantity
we want to observe. The matrix element can be rewritten in a
convenient fashion as follows. 
\[
 \langle\phi |S^{\dagger}OS|\phi\rangle =
 \bm{Tr}_{\I^+}\left[ O\rho_{red}\right],
\]
where 
\[
 \rho_{red} =
	\bm{Tr}_{H^+}\left[ S|\phi\rangle\langle\phi 
		|S^{\dagger}\right],
\]
$\bm{Tr}_{\I^+}$, $\bm{Tr}_{H^+}$ denote partial trace over
$\F(\H_{\I^+})$, $\F(\H_{H^+})$ respectively. In viewing this
expression we are lead to an interpretation that the corresponding 
final state at $\I^+$ is represented by the reduced density matrix
$\rho_{red}$. Next we generalize this argument to wider range of
initial states, which includes all mixed states. For this case an
initial state is represented not by an element of $\F_{\I^-(v>v_0)}$
but by an element of $\P$ (a density matrix on
$\F_{\I^-(v>v_0)}$). Its evolution to $\I^+$ is represented as a map 
$T$ induced by $S$ followed by the partial trace $\bm{Tr}_{H^+}$: let 
$\rho$ ($\in\P$) be an initial density matrix then the corresponding
final density matrix $T(\rho )$ ($\in\tilde{\P}$) is
\begin{equation}
 T(\rho ) = \bm{Tr}_{H^+}\left[ S\rho S^{\dagger}\right].
		\label{eqn:def-T}
\end{equation}
Note that $T$ is a linear map from $\P$ into $\tilde{\P}$.

\subsection{Thermodynamic property of $T$ }

In this subsection we show a thermal property of the map $T$, which is 
summarized as Theorem \ref{theorem:stability}. First let us calculate
a conditional probability defined as follows: 
\begin{equation}
 P(\{n_{\!_{i}\rho}\}|\{n_{\!_{i}\gamma}\}) \equiv 
     \langle \{n_{\!_{i}\rho}\}| 
     T\left( |\{n_{\!_{i}\gamma}\}\rangle
     \langle \{n_{\!_{i}\gamma}\}|\right)|\{n_{\!_{i}\rho}\}\rangle,
                \label{eqn:def-P}
\end{equation}
where
\begin{eqnarray}
 |\{n_{\!_{i}\gamma}\}\rangle & \equiv & 
      \left[\prod_{i}\frac{1}{\sqrt{n_{\!_{i}\gamma}!}}
      \left(a^{\dagger}(A\ _{i}\gamma)\right)^{n_{\!_{i}\gamma}}
      \right]|0\rangle, \nonumber\\
 |\{n_{\!_{i}\rho}\}\rangle & \equiv & 
      \left[\prod_{i}\frac{1}{\sqrt{n_{\!_{i}\rho}!}}
      \left(a^{\dagger}(\!_{i}\rho)\right)^{n_{\!_{i}\rho}}
      \right] |0\rangle.    \label{eqn:def-basis}
\end{eqnarray}
$|\{n_{\!_{i}\gamma}\}\rangle$ is a state in $\F_{\I^-(v>v_0)}$
characterized by a set of integers $n_{\!_{i}\gamma}$
($i=1,2,\cdots$) and $|\{n_{\!_{i}\rho}\}\rangle$ is a state in
$\F(\H_{\I^+})$ characterized by a set of integers $n_{\!_{i}\rho}$
($i=1,2,\cdots$). Therefore
$P(\{n_{\!_{i}\rho}\}|\{n_{\!_{i}\gamma}\})$ is a conditional
probability for a final state to be $|\{n_{\!_{i}\rho}\}\rangle$  when
the initial state is specified to $|\{n_{\!_{i}\gamma}\}\rangle$. In
the expressions $A$ is a representation of a Bogoliubov transformation
from $\H_{\I^+}\oplus\H_{H^+}$ to $\H_{\I^-}$ and $\!_{i}\gamma$ is
such a unit vector in $\H_{\I^+}\oplus\H_{H^+}$ that $A\ _{i}\gamma$
corresponds to a wave packet whose peak is located at a point on
$\I^-$ later than the formation of the horizon ($v>v_0$)
\cite{Wald}. On the other hand, $\!_{i}\rho$ is a unit vector in 
$\H_{\I^+}$ and corresponds to a wave packet on $\I^+$
\cite{Wald} (see {\it Figure} \ref{fig:background}). The probability 
(\ref{eqn:def-P}) is a generalization of $P(k|j)$ investigated by
Panangaden and Wald \cite{Panangaden&Wald} 
\footnote{
Their argument is restricted to the case when
$n_{\!_{i}\gamma}=n_{\!_{i}\rho}=0$ for all $i$ other than a
particular value.
}
It includes almost all information 
\footnote{
All the information is included in 
$T_{\{ n_{\!_{i}\rho}\} \{ n'_{\!_{i}\rho}\} }
	^{\{ n_{\!_{i}\gamma}\} \{ n'_{\!_{i}\gamma}\} }$ 
defined in Lemma \ref{lemma:off-diagonal}.
}
about a response of the scalar field on the quasi-stationary black
hole which arises from the gravitational collapse while $P(k|j)$ does
not, since any initial states on $\I^-$, which include no excitation
before the formation of the horizon ($v<v_0$.), can be represented by
using the basis $\left\{|\{n_{\!_{i}\gamma}\}\rangle\right\}$ and any
final states on $\I^+$ can be expressed by the basis
$\left\{|\{n_{\!_{i}\rho}\}\rangle\right\}$, i.e. a set of all 
$|\{n_{\!_{i}\gamma}\}\rangle$ generates $\F(\H_{\I^-(v>v_0)})$ and a
set of all $|\{n_{\!_{i}\rho}\}\rangle$ generates
$\F(\H_{\I^+})$. This is the very reason why we generalize $P(k|j)$ to 
$P(\{n_{\!_{i}\rho}\}|\{n_{\!_{i}\gamma}\})$. 

By using the S-matrix elements given in \cite{Wald}, the conditional
probability is rewritten as follows (see appendix
\ref{app:probability} for its derivation). 
\begin{eqnarray}
 P(\{ n_{\!_{i}\rho}\} |\{ n_{\!_{i}\gamma}\} )
 & = & \prod_{i}\left[ (1-x_i)x_i^{2n_{\!_{i}\rho}}
      \left( 1-|R_{i}|^{2}\right)^{n_{\!_{i}\gamma}+n_{\!_{i}\rho}}
		\right.\nonumber	\\
 & &  \times\sum_{l_{i}=0}^{\min (n_{\!_{i}\gamma},n_{\!_{i}\rho})}
      \sum_{m_{i}=0}^{\min (n_{\!_{i}\gamma},n_{\!_{i}\rho})}
      \frac{\left[ -|R_{i}|^{2}/(1-|R_{i}|^{2})
      \right]^{l_{i}+m_{i}}n_{\!_{i}\gamma}!n_{\!_{i}\rho}!}
      {l_{i}!(n_{\!_{i}\gamma}-l_{i})!(n_{\!_{i}\rho}-l_{i})!
       m_{i}!(n_{\!_{i}\gamma}-m_{i})!(n_{\!_{i}\rho}-m_{i})!} 
		\nonumber\\
 & &  \left.\times\sum_{n_i=n_{\!_{i}\rho}-\min (l_{i},m_{i})}
      ^{\infty}\frac{n_{i}!(n_{i}-n_{\!_{i}\rho}+n_{\!_{i}\gamma})!}
      {(n_{i}-n_{\!_{i}\rho}+l_{i})!(n_{i}-n_{\!_{i}\rho}+m_{i})!}
      (x_{i}^{2}|R_{i}|^{2})^{n_{i}-n_{\!_{i}\rho}}\right],
	\label{eqn:probability}
\end{eqnarray}
where $R_i$ is a reflection coefficient for the mode specified by the
integer $i$ on the Schwarzschild metric (See and $x_i$ is a 
constant defined by 
$x_{i}=\exp (-\pi (\omega_{i}-\Omega_{BH}m_{i})/\kappa )$. In the
expression, $\omega_{i}$ and $m_{i}$ are a frequency and an azimuthal
angular momentum quantum number of the mode specified by the integer
$i$, $\Omega_{BH}$ and $\kappa$ are an angular velocity and a surface
gravity of the black hole. 

Now, the expression in the squared bracket in (\ref{eqn:probability})
appears also in the calculation of $P(k|j)$. Using the result of
\cite{Panangaden&Wald}, it is easily shown that 
\begin{equation}
 P(\{ n_{\!_{i}\rho}\} |\{ n_{\!_{i}\gamma}\} ) =
      \prod_{i}\left[K_{i}\sum_{s_{i}=0}
      ^{\min (n_{\!_{i}\rho},n_{\!_{i}\gamma})}
      \frac{(n_{\!_{i}\rho}+n_{\!_{i}\gamma}-s_{i})!v_{i}^{s_{i}}}
      {s_{i}!(n_{\!_{i}\rho}-s_{i})!(n_{\!_{i}\gamma}-s_{i})!}
      \right],   \label{eqn:multi}
\end{equation}
where 
\begin{eqnarray*}
 K_{i} & = & \frac{(1-x_i)x_{i}^{2n_{\!_{i}\rho}}\left( 
      1-|R_{i}|^{2}\right)^{n_{\!_{i}\gamma}+n_{\!_{i}\rho}}}
      {\left( 1-|R_{i}|^{2}x_{i}^{2}\right)
      ^{n_{\!_{i}\gamma}+n_{\!_{i}\rho}+1}},  \\
 v_{i} & = & \frac{\left( |R_{i}|^{2}-x_{i}^{2}\right)
      \left( 1-|R_{i}|^{2}x_{i}^{2}\right)}
      {\left( 1-|R_{i}|^{2}\right)^{2}x_{i}^{2}}.
\end{eqnarray*}

This is a generalization of the result of \cite{Panangaden&Wald}, and
the following lemma is easily derived by using this expression. 

%
%
\begin{lemma}\label{lemma:balance}
For the conditional probability defined by (\ref{eqn:def-P}) the
following equality holds:
\begin{equation}
 P(\{ n_{\!_{i}\rho}=k_{i}\} |\{ n_{\!_{i}\gamma}=j_{i}\} )
      e^{-\beta_{BH}\sum_{i}j_{i}
      (\omega_{i}-\Omega_{BH}m_{i})}	=
 P(\{ n_{\!_{i}\rho}=j_{i}\} |\{ n_{\!_{i}\gamma}=k_{i}\} )
      e^{-\beta_{BH}\sum_{i}k_{i}
      (\omega_{i}-\Omega_{BH}m_{i})},\label{eqn:balance}
\end{equation}
where $\omega_{i}$ and $m_{i}$ are a frequency and an azimuthal angular
momentum quantum number of the mode specified by $i$, $\Omega_{BH}$ is 
angular velocity of the horizon and 
\[
 \beta_{BH} \equiv 2\pi /\kappa.
\]
In the expression $\kappa$ is a surface gravity of the black hole. 
\end{lemma}

Note that $\beta_{BH}^{-1}$ is the Hawking temperature of the black
hole. This lemma states that a detailed balance condition holds 
\footnote{
It guarantees that a thermal distribution of any temperature
is mapped to a thermal distribution of some other temperature closer
to the Hawking temperature, as far as the diagonal elements are
concerned.
}. 
Summing up about all $k$'s, we expect that a thermal density matrix
$\rho_{th}(\beta_{BH},\Omega_{BH})$ in $\P$ with a temperature
$\beta_{BH}^{-1}$ and a chemical potential $\Omega_{BH}$ for azimuthal 
angular momentum quantum number will be mapped by $T$ to a thermal
density matrix $\tilde{\rho}_{th}(\beta_{BH},\Omega_{BH})$ in 
$\tilde{\P}$ with the same temperature and the same chemical
potential. To show that this expectation is true, we have to prove
that all off-diagonal elements of 
$T\left(\rho_{th}(\beta_{BH},\Omega_{BH})\right)$ are zero. For this
purpose the following lemma is proved in appendix \ref{app:lemma}. 

%
%
\begin{lemma}\label{lemma:off-diagonal}
Denote a matrix element of $T$ as 
\begin{equation}
 T_{\{ n_{\!_{i}\rho}\} \{ n'_{\!_{i}\rho}\} }
	^{\{ n_{\!_{i}\gamma}\} \{ n'_{\!_{i}\gamma}\} } \equiv
 \langle\{n_{\!_{i}\rho}\}| T\left(
	|\{n_{\!_i\gamma}\}\rangle\langle\{n'_{\!_i\gamma}\}|
	\right) |\{n'_{\!_{i}\rho}\}\rangle.
\end{equation}
Then 
\begin{equation}
 T_{\{ n_{\!_{i}\rho}\} \{ n'_{\!_{i}\rho}\} }
	^{\{ n_{\!_{i}\gamma}\} \{ n'_{\!_{i}\gamma}\} } = 0, 
\end{equation}
unless
\begin{equation}
 n_{\!_{i}\gamma}-n'_{\!_{i}\gamma}= n_{\!_{i}\rho}-n'_{\!_{i}\rho}
\end{equation}
for $\forall i$.
\end{lemma}

Lemma \ref{lemma:off-diagonal} shows that all off-diagonal elements of 
$T(\rho )$ in the basis $\left\{|\{n_{\!_{i}\rho}\}\rangle\right\}$
vanish if all off-diagonal elements of $\rho$ in the basis
$\left\{|\{n_{\!_{i}\gamma}\}\rangle\right\}$ is zero. Thus, combining 
it with Lemma \ref{lemma:balance}, the following theorem is easily
proved. Note that a set of all 
$|\{n_{\!_i\gamma}\}\rangle\langle\{n'_{\!_i\gamma}\}|$ generates $\P$ 
and a set of all 
$|\{n_{\!_{i}\rho}\}\rangle\langle\{n'_{\!_{i}\rho}\}|$ generates
$\tilde{\P}$ (see the argument below (\ref{eqn:def-basis})).

%
%
\begin{theorem} \label{theorem:stability}
Consider the linear map $T$ defined by (\ref{eqn:def-T}) for a real,
massless scalar field on a background geometry which describes a
formation of a quasi-stationary black hole. Then 
\begin{equation}
 T\left(\rho_{th}(\beta_{BH},\Omega_{BH})\right)=
      \tilde{\rho}_{th}(\beta_{BH},\Omega_{BH}),
\end{equation}
where 
\begin{eqnarray}
 \rho_{th}(\beta_{BH},\Omega_{BH}) 
& \equiv &
	Z^{-1}\sum_{\{ n_{\!_{i}\gamma}\} }e^{-\beta_{BH}
	\sum_{i}n_{\!_{i}\gamma}
	\left(\omega_{i}-\Omega_{BH}m_{i}\right)}
	|\{ n_{\!_{i}\gamma}\}\rangle
	\langle\{ n_{\!_{i}\gamma}\}|,	\nonumber\\
 \tilde{\rho}_{th}(\beta_{BH},\Omega_{BH})
& \equiv &
	Z^{-1}\sum_{\{ n_{\!_{i}\rho}\} }e^{-\beta_{BH}
	\sum_{i}n_{\!_{i}\rho}
	\left(\omega_{i}-\Omega_{BH}m_{i}\right)}
	|\{ n_{\!_{i}\rho}\}\rangle
	\langle\{ n_{\!_{i}\rho}\}|,	\nonumber\\
Z
& \equiv &
	\sum_{\{ j_{i}\} }e^{-\beta_{BH}
	\sum_{i}j_{i}\left(\omega_{i}-\Omega_{BH}m_{i}\right)}.
\end{eqnarray}
\end{theorem}

$\rho_{th}(\beta_{BH},\Omega_{BH})$ and
$\tilde{\rho}_{th}(\beta_{BH},\Omega_{BH})$ can be regarded as 'grand
canonical ensemble' in $\P$ and $\tilde{\P}$ respectively, which have
a common temperature $\beta_{BH}^{-1}$ and a common chemical potential
$\Omega_{BH}$ for azimuthal angular momentum quantum number. Thus the
theorem says that the 'grand canonical ensemble' at $\I^-$ ($v>v_0$)
with special values of temperature and chemical potential evolves to a 
'grand canonical ensemble' at $\I^+$ with the same temperature and the 
same chemical potential. Note that the special values
$\beta_{BH}^{-1}$ and $\Omega_{BH}$ are determined by the background
geometry: $\beta_{BH}^{-1}$ is the Hawking temperature and
$\Omega_{BH}$ is the angular velocity of the black hole formed. This
result is used in subsection \ref{subsec:proof} to prove  the
generalized second law for the quasi-stationary black hole.


\section{The generalized second law}\label{sec:GSL}

The generalized second law is one of the most interesting conjecture
in black hole thermodynamics since it restricts ways of interaction
between a black hole and ordinary thermodynamic matter. It 
can be regarded as a generalization both of the area law of black hole
and of the second law of ordinary thermodynamics. The latter, which
states that total entropy of a system cannot decrease under a physical
evolution of a thermodynamic system, can be proved for a finite
dimensional system if a microcanonical ensemble for the system does
not change under the evolution \cite{Sorkin}.

In the previous section we have proved that the 'grand canonical
ensemble' of the scalar field does not change under the physical
evolution on a background which describes a formation of a
quasi-stationary black hole. So we expect that the generalized second
law may be proved in a way similar to the proof of the second law of
ordinary thermodynamics. For the purpose of the proof we rewrite the
generalized second law as an inequality between functionals of a
density matrix of matter fields. 

 The generalized second law of black hole thermodynamics is 
\begin{equation}
 \Delta S_{BH}+\Delta S_{matter}\geq 0,
\end{equation}
where $\Delta$ denotes a change of quantities under an evolution of
the system, $S_{BH}$ and $S_{matter}$ are black hole entropy of the
black hole and thermodynamic entropy of the matter fields
respectively. For a quasi-stationary black hole, using the first law
of the black hole thermodynamics \cite{1st-law}
\[
 \Delta S_{BH}=\beta_{BH}( \Delta M_{BH}-\Omega_{BH}\Delta J_{BH}),
\]
the conservation of total energy
\[
 \Delta M_{BH}+\Delta E_{matter}=0
\]
and the conservation of total angular momentum
\[
 \Delta J_{BH}+\Delta L_{matter}=0,
\]
it is easily shown that the generalized second law is equivalent
to the following inequality: 
\begin{equation}
 \Delta S_{matter}
 -\beta_{BH}(\Delta E_{matter}-\Omega_{BH}\Delta L_{matter})\geq 0,
\end{equation}
where $\beta_{BH}$, $\Omega_{BH}$, $M_{BH}$ and $J_{BH}$ are
inverse temperature, angular velocity, mass and angular momentum of
the black hole; $E_{matter}$ and $L_{matter}$ are energy and azimuthal 
component of angular momentum of the matter fields. Thus this is of the
form 
\begin{equation}
 U[\tilde{\rho}_0 ;\beta_{BH},\Omega_{BH})\geq 
 U[\rho_0 ;\beta_{BH},\Omega_{BH}), \label{eqn:GSL'}
\end{equation}
where $U$ is a functional of a density matrix of the matter fields
defined by 
\begin{equation}
 U[\rho ;\beta_{BH},\Omega_{BH})\equiv
      -{\bm{Tr}}\left[\rho\ln\rho\right] 
      -\beta_{BH}\left({\bm{Tr}}[{\bm{E}}\rho]
      -\Omega_{BH}{\bm{Tr}}[{\bm{L}_z}\rho]\right),
\end{equation}
$\rho_0$ and $\tilde{\rho}_0$ are an initial density matrix and the
corresponding final density matrix respectively. In the expression
$\bm{E}$ and $\bm{L}_z$ are operators corresponding to the energy and
the azimuthal component of the angular momentum. Note that
(\ref{eqn:GSL'}) is an inequality between functionals of a density
matrix of matter fields 
\footnote{
Information about the background geometry appears in the inequality as
variables which parameterize the functional.
}. 
We will prove the generalized second law by showing that this
inequality holds. Actually we do it in subsection \ref{subsec:proof}
for a quasi-stationary black hole which arises from a gravitational
collapse, using the results of Sec. \ref{sec:scalar} and a theorem
given in the following subsection. 

\subsection{Non-decreasing functional}

In this subsection a theorem which makes it possible to
construct a functional which does not decrease by a physical
evolution. It is a generalization of a result of \cite{Sorkin}. In the 
next subsection we derive (\ref{eqn:GSL'}) for a quasi-stationary black 
hole which arises from gravitational collapse, applying the theorem to
the scalar field investigated in Sec. \ref{sec:scalar}. 

Let us consider Hilbert spaces $\F$ and
$\tilde{\F}$. First we give some definitions needed for the theorem. 

%
\begin{definition}
A linear bounded operator $\rho$ on $\F$ is called a density matrix,
if it is self-adjoint, positive semi-definite and satisfies
\[
 \bm{Tr}\rho =1.
\]
\end{definition}
In the rest of this section we denote a space of all density matrices on 
$\F$ as $\P (\F )$. Evidently $\P (\F )$ is a linear convex set rather
than a linear set. 

%
\begin{definition}
A map $\T$ of $\P (\F )$ into $\P (\tilde{\F})$ is called linear, if 

\[
 \T\left( a\rho_1 +(1-a)\rho_2\right) = 
	a\T (\rho_1) + (1-a)\T (\rho_2)
\]
for $0\leq\!^\forall a\leq 1$ and $\!^\forall\rho_1$, 
$\!^\forall\rho_2$ $(\in\P (\F ))$.
\end{definition}
By this definition it is easily proved by induction that 
\begin{equation}
 \T\left(\sum_{i=1}^N a_i\rho_i\right) =
	\sum_{i=1}^N a_i\T (\rho_i ),	\label{eqn:linearity}
\end{equation}
if $a_i\geq 0$, $\sum_{i=1}^N a_i =1$ and $\rho_i\in\P (\F )$.

Now we prove the following lemma which concerns the $N\to\infty$ limit
of the left hand side of (\ref{eqn:linearity}). We use this lemma in
the proof of theorem \ref{theorem:general}.
%
\begin{lemma}	\label{lemma:WOT-limit}
Consider a linear map $\T$ of $\P (\F )$ into $\P (\tilde{\F})$ and an
element $\rho_0$ of $\P (\F )$. For a diagonal decomposition
\[
 \rho_0 = \sum_{i=1}^\infty p_i|i\rangle\langle i|,
\]
define a series of density matrices of the form 
\begin{equation}
 \rho_n = \sum_{i=1}^n p_i/a_n|i\rangle\langle i|
	\qquad (n=N,N+1,\cdots ),
\end{equation}
where 
\[
 a_n \equiv \sum_{i=1}^n p_i
\]
and $N$ is large enough that $a_N >0$.
Then 
\begin{equation}
 \lim_{n\to\infty}\langle\Phi |\T(\rho_n)|\Psi\rangle =
	\langle\Phi |\T(\rho_0)|\Psi\rangle
\end{equation}
for arbitrary elements $|\Phi\rangle$ and $|\Psi\rangle$ of
$\tilde{\F}$. 
\end{lemma}

This lemma says that $\T (\rho_n)$ has a weak-operator-topology-limit
$\T (\rho_0)$.

%
%
\begin{proof}
By definition,
\begin{equation}
 \rho_0 = a_n\rho_n + (1-a_n)\rho'_n,
\end{equation}
where
\[
 \rho'_n = \left\{ \begin{array}{lc}
	\sum_{i=n+1}^\infty p_i/(1-a_n)|i\rangle\langle i| &
	(a_n <1)	\\
	\rho_n	&
	(a_n =1)
 \end{array}\right..
\]
Then the linearity of $\T$ shows
\[
 \langle\Phi |\T(\rho_0)|\Psi\rangle =
	a_n\langle\Phi |\T(\rho_n)|\Psi\rangle +
	(1-a_n)\langle\Phi |\T(\rho'_n)|\Psi\rangle.
\]
Thus, if $\langle\Phi |\T(\rho'_n)|\Psi\rangle$ is finite in
$n\to\infty$ limit, then the lemma is established since 
\[
 \lim_{n\to\infty}a_n=1.
\]
For the purpose of proving the finiteness of 
$\langle\Phi |\T(\rho'_n)|\Psi\rangle$, it is sufficient to show that
$|\langle\Phi |\tilde{\rho}|\Psi\rangle|$ is bounded from above by 
$\|\Phi\|\ \|\Psi\|$ for an arbitrary element $\tilde{\rho}$ of 
$\P (\tilde{\F})$. This is easy to prove as follows. 
\begin{equation}
 |\langle\Phi |\tilde{\rho}|\Psi\rangle | = 
 |\sum_i\tilde{p}_i\langle\Phi |\tilde{i}\rangle
	\langle \tilde{i}|\Psi\rangle| \leq 
 \sum_i|\langle\Phi |\tilde{i}\rangle\langle\tilde{i}|\Psi\rangle |
 \leq \|\Phi\|\ \|\Psi\| ,
\end{equation}
where we have used a diagonal decomposition
\[
 \tilde{\rho} = \sum_i\tilde{p}_i|\tilde{i}\rangle\langle\tilde{i}|.
\]
\end{proof}
\QED

%
\begin{theorem} \label{theorem:general}
Assume the following three assumptions: 
{\bm a}. $\T$ is a linear map of $\P (\F )$ into $\P (\tilde{\F})$,
$\quad$ 
{\bm b}. $f$ is a continuous function convex to below and there are
non-negative constants $c_1$, $c_2$ and $c_3$ such that 
$|f((1-\epsilon)x)-f(x)|\leq |\epsilon|(c_1|f(x)|+c_2|x|+c_3)$ for
$\!^\forall x$ $(\geq 0)$ and sufficiently small $|\epsilon |$,\quad 
{\bm c}. there are positive definite density matrices $\rho_\infty$
$(\in\P (\F ))$ and $\tilde{\rho}_\infty$ $(\in\P (\tilde{\F}))$ such
that $\T (\rho_\infty )=\tilde{\rho}_\infty$.

If $[\rho_\infty,\rho_0 ]=[\tilde{\rho}_\infty,\T (\rho_0)]=0$ and
$\bm{Tr}[\rho_\infty |f(\rho_0\rho^{-1}_\infty )|]<\infty$, then 
\begin{equation}
  \tilde{\U}\left[\T (\rho_{0})\right]\geq \U\left[\rho_{0}\right],
	\label{eqn:theorem}
\end{equation}
where
\begin{eqnarray}
 \U\left[\rho\right] & \equiv & -{\bf Tr}\left[\rho_{\infty}f\left(\rho
      \rho_{\infty}^{-1}\right)\right],\nonumber\\
 \tilde{\U}\left[\tilde{\rho}\right] & \equiv & 
      -{\bf Tr}\left[\tilde{\rho}_{\infty}f\left(\tilde{\rho}
      \tilde{\rho}_{\infty}^{-1}\right)\right].
\end{eqnarray}
\end{theorem}
As stated in the first paragraph of this subsection, theorem
\ref{theorem:general} is used in subsection \ref{subsec:proof} to
prove the generalized second law for a quasi-stationary black hole
which arises from gravitational collapse. 

%
%
\begin{proof}
First let us decompose the density matrices diagonally as follows:
\begin{eqnarray}
 \rho_0 
& = & \sum_{i=1}^\infty p_i|i\rangle\langle i|,\quad
 \rho_\infty
 =  \sum_{i=1}^\infty q_i|i\rangle\langle i|,	\nonumber\\
 \T (\rho_0)
& = & \sum_{i=1}^\infty\tilde{p}_i|\tilde{i}\rangle\langle\tilde{i}|,
	\quad
 \T (\rho_\infty )
 =  \sum_{i=1}^\infty\tilde{q}_i|\tilde{i}\rangle\langle\tilde{i}|.
\end{eqnarray}
Then by lemma \ref{lemma:WOT-limit} and (\ref{eqn:linearity}),
\begin{equation}
 \tilde{p}_i 
 = \langle\tilde{i}|\T (\rho_0)|\tilde{i}\rangle
 = \lim_{n\to\infty}\sum_{j=1}^n A_{ij}p_j/a_n,	\label{eqn:tilde-p}
\end{equation}
where $a_n\equiv\sum_{i=1}^np_i$ and 
$A_{ij}\equiv\langle\tilde{i}|\T 
	(|j\rangle\langle j|)|\tilde{i}\rangle$. $A_{ij}$ has the
following properties:
\[
 \sum_{i=1}^\infty A_{ij} =1,\quad 0\leq A_{ij}\leq 1.
\]
Similarly it is shown that
\[
 \tilde{q}_i
= \lim_{n\to\infty}\sum_{j=1}^n A_{ij}q_j/b_n,
\]
where $b_n\equiv\sum_{i=1}^n q_i$. By (\ref{eqn:tilde-p}) and the
continuity of $f$, it is shown that 
\begin{equation}
 f(\tilde{p}_i/\tilde{q}_i) = \lim_{n\to\infty}
	f\left(\sum_{j=1}^n A_{ij}\frac{p_j/a_n}{\tilde{q}_i}\right).
\end{equation}

Next define $C^n_i$ and $\tilde{C}^n_i$ by 
\begin{equation}
 C^n_i \equiv \sum_{j=1}^n A_{ij}q_j/\tilde{q}_i,\quad
 \tilde{C^n_i} \equiv C^n_i/a_n,
\end{equation}
then the convex property of $f$ means
\[
 f(\tilde{p}_i/\tilde{q}_i) \leq
	\lim_{n\to\infty}\sum_{j=1}^n\frac{A_{ij}q_j}{C^n_i\tilde{q}_i}
	f(\tilde{C}^n_ip_j/q_j)
\]
since
\[
 \sum_{j=1}^n\frac{A_{ij}q_j}{C^n_i\tilde{q}_i} = 1,
	\frac{A_{ij}q_j}{C^n_i\tilde{q}_i} \geq 0.
\]
Hence
\begin{equation}
 -\tilde{\U}[\T (\rho_0)] 
 = \sum_{i=1}^\infty\tilde{q}_if(\tilde{p}_i/\tilde{q}_i)
 \leq \sum_{i=1}^\infty\lim_{n\to\infty}\sum_{j=1}^n
	\frac{A_{ij}q_j}{C^n_i}
	f(\tilde{C}^n_ip_j/q_j).
\end{equation}

Since $C^n_i$ and $\tilde{C^n_i}$ satisfy 
\[
 \lim_{n\to\infty}C^n_i =\lim_{n\to\infty}\tilde{C^n_i} = 1,
\]
it is implied by the assumption about $f$ that 
\[
 \left| f\left(\frac{\tilde{C}^n_ip_j}{q_j}\right) -
	f\left(\frac{p_j}{q_j}\right)\right|
 \leq |1-\tilde{C}^n_i|\left( c_1|f(p_j/q_j)|+c_2p_j/q_j+c_3\right)
\]
for sufficiently large $n$. Therefore
\begin{eqnarray*}
 \left|\sum_{j=1}^n\frac{A_{ij}q_j}{C^n_i}
	\left( f(\tilde{C}^n_ip_j/q_j) - f(p_j/q_j)\right)\right|
& \leq &
 \frac{|1-\tilde{C}^n_i|}{C^n_i}
	\left(c_1\sum_{j=1}^nA_{ij}q_j|f(p_j/q_j)| + 
	c_2\sum_{j=1}^nA_{ij}p_j + 
	c_3\sum_{j=1}^nA_{ij}q_j\right)	\\
& \leq &
 \frac{|1-\tilde{C}^n_i|}{C^n_i}
	\left(c_1\sum_{j=1}^nq_j|f(p_j/q_j)| + 
	c_2\sum_{j=1}^np_j +
	c_3\sum_{j=1}^nq_j\right),
\end{eqnarray*}
where we have used $0\leq A_{ij}\leq 1$ to obtain the last
inequality. Since the first term in the brace in the last expression
is finite in $n\to\infty$ limit by the assumption of the absolute
convergence of $\U\left[\rho_0\right]$ and all the other terms in the
brace are finite,
\[
 \lim_{n\to\infty}\left|\sum_{j=1}^n\frac{A_{ij}q_j}{C^n_i}
	\left( f(\tilde{C}^n_ip_j/q_j) - f(p_j/q_j)\right)\right|
 = 0.
\]
Moreover, by the absolute convergence of $\U\left[\rho_0\right]$, it
is easily shown that
\[
 \lim_{n\to\infty}\left|\left(\frac{1}{C^n_i}-1\right)
	\sum_{j=1}^n A_{ij}q_jf(p_j/q_j)\right|=0.
\]
Thus
\begin{equation}
 -\tilde{\U}[\T (\rho_0 )]\leq
	\sum_{i=1}^\infty\sum_{j=1}^\infty A_{ij}q_jf(p_j/q_j).
		\label{eqn:theorem'}
\end{equation}
We can interchange sum over $i$ and sum over $j$ in the right hand
side of (\ref{eqn:theorem'}) since it converges absolutely by the
absolute convergence of $\U\left[\rho_0\right]$. Hence 
\[
 -\tilde{\U}[\T (\rho_0 )]\leq
	\sum_{j=1}^\infty q_jf(p_j/q_j) = -\U [\rho_0].
\]
\end{proof}
\QED

\subsection{Proof of the generalized second law}
	\label{subsec:proof}

Let us combine theorem \ref{theorem:stability} with theorem
\ref{theorem:general} to prove the generalized second law. In theorem
\ref{theorem:general} set the linear map $\T$, the convex function
$f(x)$ and the density matrices $\rho_{\infty}$ and
$\tilde{\rho}_{\infty}$ as follows. 
\begin{eqnarray}
 \T & = & T,\nonumber\\
 f(x) & = & x\ln x,\nonumber\\
 \rho_{\infty} & = & \rho_{th}(\beta_{BH},\Omega_{BH}),\nonumber\\
 \tilde{\rho}_{\infty} & = & 
	\tilde{\rho}_{th}(\beta_{BH},\Omega_{BH}).
\end{eqnarray}
Note that it is theorem \ref{theorem:stability} that makes such a
setting possible. Hence, if an initial state $\rho_{0}$ and the
corresponding final state $T(\rho_{0})$ satisfy 
\begin{equation}
 \left[\rho_{0},\rho_{th}(\beta_{BH},\Omega_{BH})\right]=
 \left[T(\rho_{0}),\rho_{th}(\beta_{BH},\Omega_{BH})\right]=0
   \label{eqn:initial-cond}
\end{equation}
and $\U [\rho_0]$ converges absolutely, theorem 
\ref{theorem:general} can be applied to the system of the 
quasi-stationary black hole and the scalar field around it. Now
\begin{eqnarray}
 \U [\rho_0] & = & -\bm{Tr}\left[\rho_0\ln\rho_0\right]
	-\beta_{BH}\left(\bm{Tr}\left[\bm{E}\rho_0\right]
		-\Omega_{BH}\bm{Tr}\left[\bm{L}_z\rho_0\right]
			\right) -\ln Z	\nonumber\\
 & = & U\left[\rho_{0};\beta_{BH},\Omega_{BH}\right) -\ln Z,
					\nonumber\\
 \tilde{\U}[\T (\tilde{\rho}_0 )]& = & 
-\bm{Tr}\left[\tilde{\rho}_0\ln\tilde{\rho}_0\right]
-\beta_{BH}\left(\bm{Tr}\left[\tilde{\bm{E}}\tilde{\rho}_0\right]
-\Omega_{BH}\bm{Tr}\left[\tilde{\bm{L}}_z\tilde{\rho}_0\right]
			\right) -\ln Z	\nonumber\\
 & = & U\left[ T(\tilde{\rho}_{0});\beta_{BH},\Omega_{BH}\right) 
	-\ln Z,
\end{eqnarray}
where
\begin{eqnarray*}
 \bm{E} & \equiv & \sum_{\{n_{\!_{i}\gamma}\} }
	\left(\sum_i n_{\!_{i}\gamma}\omega_i\right)
	|\{n_{\!_{i}\gamma}\}\rangle\langle\{n_{\!_{i}\gamma}\}|,\\ 
 \bm{L}_z & \equiv & \sum_{\{n_{\!_{i}\gamma}\} }
	\left(\sum_i n_{\!_{i}\gamma}m_i\right)
	|\{n_{\!_{i}\gamma}\}\rangle\langle\{n_{\!_{i}\gamma}\}|,
\end{eqnarray*}
and 
\begin{eqnarray*}
 \tilde{\bm{E}} & \equiv & \sum_{\{n_{\!_{i}\rho}\} }
	\left(\sum_i n_{\!_{i}\rho}\omega_i\right)
	|\{n_{\!_{i}\rho}\}\rangle\langle\{n_{\!_{i}\rho}\}|,\\ 
 \tilde{\bm{L}}_z & \equiv & \sum_{\{n_{\!_{i}\rho}\} }
	\left(\sum_i n_{\!_{i}\rho}m_i\right)
	|\{n_{\!_{i}\rho}\}\rangle\langle\{n_{\!_{i}\rho}\}|.
\end{eqnarray*}
Thus the inequality (\ref{eqn:theorem}) in this case is
(\ref{eqn:GSL'}) itself, which in turn is equivalent to the
generalized second law. Finally, theorem \ref{theorem:general} proves
the generalized second law for a quasi-stationary black hole which
arises from gravitational collapse, provided that an initial density
matrix $\rho_0$ of the scalar field satisfies the above
assumptions. For example, it is guaranteed by lemma
\ref{lemma:off-diagonal} that if $\rho_{0}$ is diagonal in the
basis $\{ |\{n_{\!_{i}\gamma}\}\rangle\}$ then $T(\rho_{0})$ is also
diagonal in the basis $\{ |\{n_{\!_{i}\rho}\}\rangle\}$ and
(\ref{eqn:initial-cond}) is satisfied. The assumption of the absolute 
convergence of $U[\rho_{0};\beta_{BH},\Omega_{BH})$ holds whenever
initial state $\rho_0$ at $\I^-$ contains at most finite number of 
excitations. Note that although $\rho_{th}(\beta_{BH},\Omega_{BH})$
contains infinite number of excitations by definition, $\rho_0$ has
not to do. Therefore the assumptions are satisfied when $\rho_0$ is
diagonal in the basis $\{ |\{n_{\!_{i}\gamma}\}\rangle\}$ and contains
at most finite number of excitations.


\section{Summary and discussion}\label{sec:summary}

In summary we have proved the generalized second law for a
quasi-stationary black hole which arises from gravitational
collapse. To prove it we have derived the thermal property of the
semiclassical evolution of a real massless scalar
field on the quasi-stationary black hole background and have given a
method for searching a non-decreasing functional. These are
generalizations of results of \cite{Panangaden&Wald} and \cite{Sorkin}
respectively.

\begin{figure}
\centerline{\epsfbox{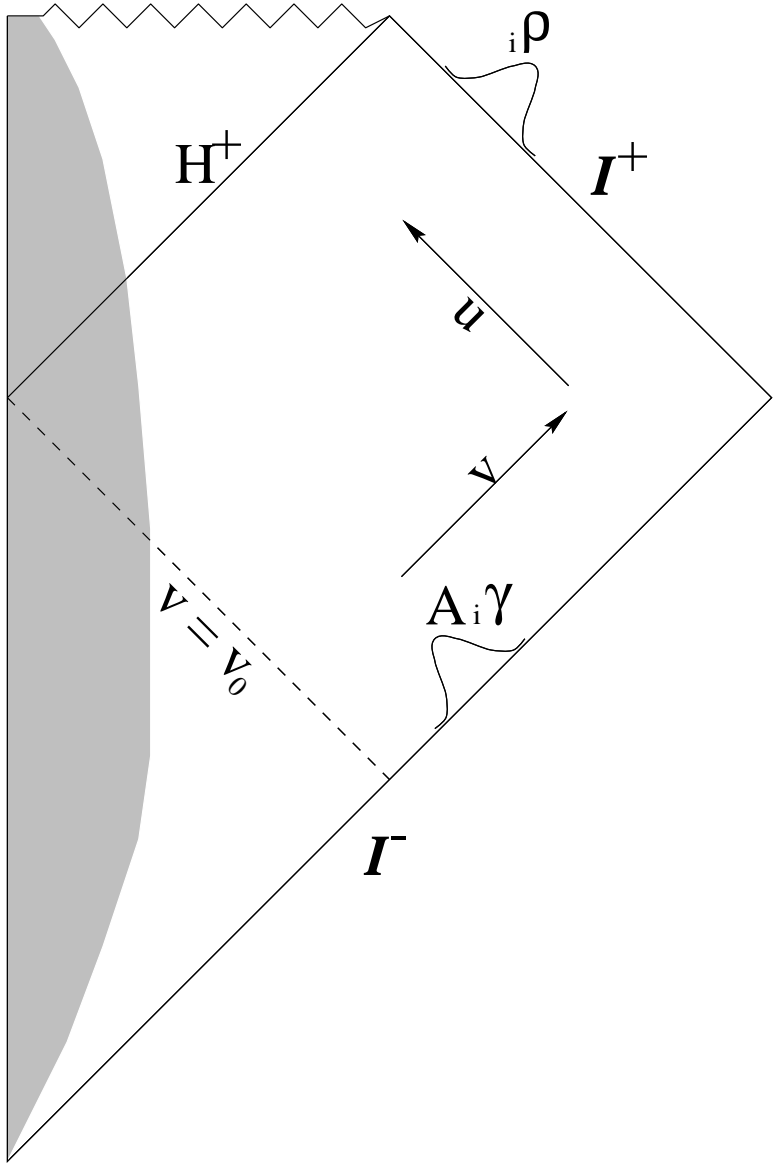}\vspace{1cm}}
 \caption{
A conformal diagram of a background geometry which describes a 
gravitational collapse. $\I^-$ and $\I^+$ are the past null infinity
and the future null infinity, respectively and $H^+$ is the future
event horizon. \\
Shaded region represents collapsing materials which forms the black
hole. Besides the collapsing matter, we consider a real massless
scalar field and investigate a scattering problem by the black hole
after its formation ($v>v_0$). Thus we specify possible initial states 
at $\I^-$ to those states which are excited from the vacuum by only
modes whose support is within $v>v_0$ (elements of
$\F_{\I^-(v>v_0)}$),
and possible mixed states constructed from them (elements of $\P$). In
the diagram, $A\ _i\gamma$ ($i=1,2,\cdots$) is a mode function
corresponding to a wave packet whose peak is at $v>v_0$ on $\I^-$, 
$\!_i\rho$ ($i=1,2,\cdots$) is a mode function corresponding to a wave 
packet on $\I^+$.
}
\label{fig:background}
\end{figure}

Now we make a comments on Frolov and Page's statement that their proof 
of the generalized second law may be applied to the case of the black
hole formed by a gravitational collapse \cite{Frolov&Page}. Their proof 
for a quasi-stationary eternal black hole is based on the following
two assumptions: (1) a state of matter fields on the past horizon is 
thermal one; (2) a set of radiation modes on the past horizon and one
on the past null infinity are quantum mechanically uncorrelated. These
two assumptions are reasonable for the eternal case since a black hole
emit a thermal radiation. In the case of a black hole which arises
from a gravitational collapse, we might expect that things would go
well by simply replacing the past horizon with a null surface at a
moment of a formation of a horizon ($v=v_0$ surface in {\it
Figure \ref{fig:background}}). However, a state of the matter fields on
the past horizon is completely determined by a state of the fields
before the horizon formation ($v<v_0$ in {\it Figure}
\ref{fig:background}), in which there is no causal effect of the
existence of future horizon. Since the essential origin of the thermal
radiation from a black hole is the existence of a horizon, the state
of the fields on the null surface has not to be a thermal one. Hence
the assumption (1) becomes ill in this case. Although the above
replacement may be the most extreme one, an intermediate replacement
causes an intermediate violation of the assumption (1) due to the
correlation between modes on the future null infinity and modes on the
past null infinity located after the horizon formation. The
correlation can be seen in (\ref{eqn:multi}) explicitly. Thus we
conclude that their proof can not be applied to the case of the black
hole formed by a gravitational collapse.  

Finally we discuss a generalization of our proof to a dynamical
background. For the case of a dynamical background $\beta_{BH}$ and
$\Omega_{BH}$ are changed from time to time by a possible
backreaction. Thus, to prove the generalized second law for the
dynamical background, we have to generalize theorem
\ref{theorem:stability} to the dynamical case consistently with the 
backreaction. Once this can be achieved, theorem \ref{theorem:general} 
seems useful to prove the generalized second law for the dynamical
background.

\vspace{1cm}
\centerline{\bf Acknowledgment}
The author thanks Prof. H. Kodama for continuous encouragement and
discussions. He also thanks Dr. M. Seriu and Dr. M. Siino for
helpful discussions, and Dr. T. Chiba for his encouragement for him to
write this manuscript.

\appendix

\section{The conditional probability}	\label{app:probability}

In this appendix we reduce (\ref{eqn:def-P}) to
(\ref{eqn:probability}). First the $S$-matrix obtained by
\cite{Wald} is
\begin{eqnarray}
 S|0\rangle & = & 
      N\sum_{n=0}^{\infty}\frac{\sqrt{(2n)!}}{2^{n}n!}
      \left(\stackrel{n}{\otimes}\epsilon\right)_{sym},\nonumber\\
 Sa^{\dagger}(A\ _{i}\gamma)S^{-1} & = & 
      R_ia^{\dagger}(\!_{i}\rho)+T_ia^{\dagger}(\!_{i}\sigma),
\end{eqnarray}
where $\epsilon$ and $N$ are a bivector and a normalization constant
defined by 
\[
 \epsilon = 2\sum_{i}x_{i}(\!_{i}\lambda\otimes\!_{i}\tau)_{sym},
	\quad N=\prod_i\sqrt{1-x_i},
\]
where
\[
 x_{i} = \exp\left( -\pi
	(\omega_{i}-\Omega_{BH}m_{i})/\kappa\right).
\]
In the expression, $\omega_{i}$ and $m_{i}$
are a frequency and an azimuthal angular momentum quantum number of a
mode specified by integer $i$, $\Omega_{BH}$ and $\kappa$ are an
angular velocity and a surface gravity of the black
hole. $\!_{i}\gamma$, $\!_{i}\rho$, $\!_{i}\sigma$, $\!_{i}\lambda$
and $\!_{i}\tau$ are unit vectors in $\H_{\I^+}\oplus\H_{H^+}$ defined
in \cite{Wald}, and the former four are related as follows:
\begin{eqnarray}
 \!_{i}\gamma^{a}  & = & T_{i}\ _{i}\sigma^{a}
                         +R_{i}\ _{i}\rho^{a},	\nonumber\\
 \!_{i}\lambda^{a} & = & t_{i}\ _{i}\rho^{a}
                         +r_{i}\ _{i}\sigma^{a},	
			\label{eqn:gamma-lambda}
\end{eqnarray}
where $t_{i}$, $T_{i}$ are transmission coefficients for the mode
specified by the integer $i$ on the Schwarzschild metric \cite{Wald}
and $r_{i}$, $R_{i}$ are reflection coefficients. They satisfy 
\footnote{The last two equations are consequences of the time
reflection symmetry of the Schwarzschild metric.}
\begin{eqnarray}
 |t_{i}|^{2}+|r_{i}|^{2} & = & |T_{i}|^{2}+|R_{i}|^{2}=1, 
				\nonumber\\
 t_{i} = T_{i}, & & r_{i}=-R_{i}^{*}T_{i}/T_{i}^{*}.\label{eqn:trTR}
\end{eqnarray}

By using the S-matrix, we obtain
\begin{eqnarray}
 S|\{ n_{\!_{i}\gamma}\}\rangle & = & 
      N\left[\prod_{i}\frac{1}{\sqrt{n_{\!_{i}\gamma}!}}\left[ 
      R_{i}a^{\dagger}(\!_{i}\rho )+T_{i}a^{\dagger}(\!_{i}\sigma )
      \right]^{n_{\!_{i}\gamma}}\right]\sum_{n=0}^{\infty}
      \frac{\sqrt{(2n)!}}{2^{n}n!}
      \left(\stackrel{n}{\otimes}\epsilon\right)_{sym}\nonumber\\
 & = & N\sum_{n=0}^{\infty}{\sum}'\left[\prod_{i}
      \frac{1}{\sqrt{n_{\!_{i}\gamma}!}}\left(\begin{array}{c}
      n_{\!_{i}\gamma}\\m_{i}\end{array}\right)R_{i}^{m_{i}}
      T_{i}^{n_{\!_{i}\gamma}-m_{i}}\right]
      \sqrt{(2n)!}\left[\prod_{i}\frac{x_{i}^{n_{i}}}{n_i!}
      \left(\begin{array}{c}n_{i}\\l_{i}\end{array}\right)t_i^{l_{i}}
      r_i^{n_{i}-l_{i}}\right]  \nonumber \\
 & &  \times\sqrt{\frac{(2n+\sum_{i}n_{\!_{i}\gamma})!}
      {(2n)!}}\left(\prod_{i}\stackrel{n_{i}}{\otimes}\!_{i}\tau
      \stackrel{l_{i}+m_{i}}{\otimes}\!_{i}\rho
      \stackrel{n_{i}-l_{i}+n_{\!_{i}\gamma}-m_{i}}{\otimes}
      \!_{i}\sigma\right)_{sim} \nonumber\\
 & = & N\sum_{n_{i}=0}^{\infty}
      \sum_{m_{i}=0}^{n_{\!_{i}\gamma}}\sum_{l_{i}=0}^{n_{i}}
      \sqrt{\left(\sum_{i}(2n_{i}+n_{\!_{i}\gamma})\right) !}
      \prod_{i}\left[\frac{1}
      {\sqrt{n_{\!_{i}\gamma}!}}\cdot\frac{x_{i}^{n_{i}}}{n_{i}!}
      \left(\begin{array}{c}n_{\!_{i}\gamma}\\m_{i}\end{array}\right)
      \left(\begin{array}{c}n_{i}\\l_{i}\end{array}\right)
      R_{i}^{m_{i}}T_{i}^{n_{\!_{i}\gamma}-m_{i}}t_{i}^{l_{i}}r_{i}
      ^{n_{i}-l_{i}}\right]   \nonumber  \\
 & &  \times\left(\prod_{i}\stackrel{n_{i}}{\otimes}
      \!_{i}\tau\stackrel{l_{i}+m_{i}}{\otimes}\!_{i}\rho
      \stackrel{n_{i}-l_{i}+n_{\!_{i}\gamma}-m_{i}}{\otimes}
      \!_{i}\sigma\right)_{sym},	\label{eqn:S|n>}
\end{eqnarray}
where ${\sum}'$ denotes a summation with respect to $n_i$, $m_i$ and
$l_i$ over the following range: $\sum_in_i=n$, $n_i\geq 0$, 
$0\leq m_i\leq n_{\!_i\gamma}$, $0\leq l_i\leq n_i$. In
(\ref{eqn:S|n>}), those orthonormal basis vectors in 
$|\{ n_{\!_{i}\rho}\}\rangle\otimes\F (\H_{H^+})$ that have a
non-vanishing inner product with $S|\{ n_{\!_{i}\gamma}\}\rangle$
appear in the form 
\footnote{The number of the 'particle' $\!_i\sigma$ in
(\ref{eqn:S|n>}) is $n_{i}+n_{\!_{i}\gamma}-n_{\!_{i}\rho}$, setting
the number of the 'particle' $\!_i\rho$ to $n_{\!_{i}\rho}$.}
\begin{equation}
 \sqrt{\frac{\left(\sum_{i}(2n_{i}+n_{\!_{i}\gamma})\right) !}
      {\prod_{i}[n_{i}!n_{\!_{i}\rho}!(n_{i}+n_{\!_{i}\gamma}
      -n_{\!_{i}\rho})!]}}\left(\prod_{i}\stackrel{n_{i}}{\otimes}
      \!_{i}\tau\stackrel{n_{\!_{i}\rho}}{\otimes}\!_{i}\rho
      \stackrel{n_{i}+n_{\!_{i}\gamma}-n_{\!_{i}\rho}}
      {\otimes}\!_{i}\sigma\right)_{sym}.\label{non-zero-term}
\end{equation}
Thus, when we calculate
$\left(\langle\{ n_{\!_{i}\rho}\} |\otimes\langle H|\right)S 
	|\{ n_{\!_i\gamma}\}\rangle$, 
the summation in (\ref{eqn:S|n>}) is reduced to a summation with
respect to $n_i$ and $m_i$ over the range 
$n_i\geq\max (0,n_{\!_{i}\rho}-n_{\!_{i}\gamma})$,  
$\max (0,n_{\!_{i}\rho}-n_{i})\leq m_{i}\leq
	\min(n_{\!_{i}\rho},n_{\!_{i}\gamma})$ 
with $l_i=n_{\!_{i}\rho}-m_i$ 
\footnote{
The range is obtained by inequalities $n_i\geq 0$, 
$0\leq m_i\leq n_{\!_{i}\gamma}$, $0\leq l_i\leq n_i$, 
$l_i+m_i=n_{\!_{i}\rho}$ and 
$n_i+n_{\!_{i}\gamma}-n_{\!_{i}\rho}\geq 0$.
}.
Here $|H\rangle$ is an element of $\F(\H_{H^+})$. Paying attention to
this fact, we can obtain the following expression of the conditional
probability.
\begin{eqnarray}
 P(\{ n_{\!_{i}\rho}\} |\{ n_{\!_{i}\gamma}\} )
 & = & |N|^{2}
	\sum_{\{ n_{i}\geq\max (0,n_{\!_{i}\rho}-n_{\!_{i}\gamma})\}}
      \left(\sum_{i}(2n_{i}+n_{\!_{i}\gamma})\right) ! \nonumber\\
 & &  \times\prod_{i}\left[\frac{x_{i}^{2n_{i}}}
      {n_{\!_{i}\gamma}!(n_{i}!)^{2}}
      \left|\sum
      _{m_{i}=\max (0,n_{\!_{i}\rho}-n_{i})}
      ^{\min (n_{\!_{i}\rho},n_{\!_{i}\gamma})}
      \left(\begin{array}{c}n_{\!_{i}\gamma}\\m_{i}\end{array}\right)
      \left(\begin{array}{c}n_{i}\\n_{\!_{i}\rho}-m_{i}\end{array}\right)
      R_{i}^{m_{i}}T_{i}^{n_{\!_{i}\gamma}-m_{i}}
      t_{i}^{n_{\!_{i}\rho}-m_{i}}r_{i}^{n_{i}-n_{\!_{i}\rho}+m_{i}}
      \right|^{2}\right]   \nonumber\\
 & &  \times\left|\langle\sqrt{
      \frac{\left(\sum_{i}(2n_{i}+n_{\!_{i}\gamma})\right) !}
      {\prod_{i}\left[n_{i}!
      n_{\!_{i}\rho}!(n_{i}+n_{\!_{i}\gamma}-n_{\!_{i}\rho})!\right]}}
      \prod_{i}\left(\stackrel{n_{i}}{\otimes}
      \!_{i}\tau\stackrel{n_{\!_{i}\rho}}{\otimes}\!_{i}\rho
      \stackrel{n_{i}+n_{\!_{i}\gamma}-n_{\!_{i}\rho}}
      {\otimes}\!_{i}\sigma\right)_{sym},\right.\nonumber\\
 & &  \left.\hspace{7cm}
      \prod_{i}\left(\stackrel{n_{i}}{\otimes}
      \!_{i}\tau\stackrel{n_{\!_{i}\rho}}{\otimes}\!_{i}\rho
      \stackrel{n_{i}+n_{\!_{i}\gamma}-n_{\!_{i}\rho}}
      {\otimes}\!_{i}\sigma\right)_{sym}\rangle\right|^{2}.
\end{eqnarray}
The inner product in the last expression equals to 
\footnote{(\ref{non-zero-term}) is normalized to have unit norm.}
\[
 \sqrt{\frac{\prod_{i}\left[ 
      n_{i}!n_{\!_{i}\rho}!(n_{i}+n_{\!_{i}\gamma}-n_{\!_{i}\rho})!\right]}
      { \left(\sum_{i}(2n_{i}+n_{\!_{i}\gamma})\right) !}}.
\]
Finally, by using (\ref{eqn:trTR}) and exchanging the order of the
summation suitably, we can obtain
\begin{eqnarray*}
 P(\{ n_{\!_{i}\rho}\} |\{ n_{\!_{i}\gamma}\} )
 & = & \prod_{i}\left[ (1-x_i)x_i^{2n_{\!_{i}\rho}}
      \left( 1-|R_{i}|^{2}\right)^{n_{\!_{i}\gamma}+n_{\!_{i}\rho}}
		\right.\nonumber	\\
 & &  \times\sum_{l_{i}=0}^{\min (n_{\!_{i}\gamma},n_{\!_{i}\rho})}
      \sum_{m_{i}=0}^{\min (n_{\!_{i}\gamma},n_{\!_{i}\rho})}
      \frac{\left[ -|R_{i}|^{2}/(1-|R_{i}|^{2})
      \right]^{l_{i}+m_{i}}n_{\!_{i}\gamma}!n_{\!_{i}\rho}!}
      {l_{i}!(n_{\!_{i}\gamma}-l_{i})!(n_{\!_{i}\rho}-l_{i})!
       m_{i}!(n_{\!_{i}\gamma}-m_{i})!(n_{\!_{i}\rho}-m_{i})!} 
		\nonumber\\
 & &  \left.\times\sum_{n_i=n_{\!_{i}\rho}-\min (l_{i},m_{i})}
      ^{\infty}\frac{n_{i}!(n_{i}-n_{\!_{i}\rho}+n_{\!_{i}\gamma})!}
      {(n_{i}-n_{\!_{i}\rho}+l_{i})!(n_{i}-n_{\!_{i}\rho}+m_{i})!}
      (x_{i}^{2}|R_{i}|^{2})^{n_{i}-n_{\!_{i}\rho}}\right].
\end{eqnarray*}
This is what we have to show.

\section{A proof of Lemma 2}	\label{app:lemma}
In this appendix we give a proof of Lemma \ref{lemma:off-diagonal}.
%
%
\begin{proof}
Since a set of all $\!_i\tau$ and $\!_i\sigma$ generates $\H_{H^+}$
\cite{Wald}, the definition of 
$T_{\{ n_{\!_{i}\rho}\} \{ n'_{\!_{i}\rho}\} }
^{\{ n_{\!_{i}\gamma}\} \{ n'_{\!_{i}\gamma}\} }$ leads
\begin{equation}
  T_{\{ n_{\!_{i}\rho}\} \{ n'_{\!_{i}\rho}\} }
	^{\{ n_{\!_{i}\gamma}\} \{ n'_{\!_{i}\gamma}\} } =
 \sum_{\{ n_{\!_i\sigma}\},\{ n_{\!_i\tau}\}}
	\langle\{n_{\!_i\tau},n_{\!_i\rho},n_{\!_i\sigma}\}|
	S|\{n_{\!_i\gamma}\}\rangle
	\langle\{n'_{\!_i\gamma}\}|
	S|\{n_{\!_i\tau},n'_{\!_i\rho},n_{\!_i\sigma}\}\rangle,
\end{equation}
where
\[
 |\{n_{\!_i\tau},n_{\!_i\rho},n_{\!_i\sigma}\}\rangle	\equiv
	\prod_i\left[\frac{1}
	{\sqrt{n_{\!_i\tau}!n_{\!_i\rho}!n_{\!_i\sigma}!}}
	\left(a^{\dagger}(\!_{i}\tau)\right)^{n_{\!_{i}\tau}}
	\left(a^{\dagger}(\!_{i}\rho)\right)^{n_{\!_{i}\rho}}
	\left(a^{\dagger}(\!_{i}\sigma)\right)^{n_{\!_{i}\sigma}}
	\right] |0\rangle.
\]
In the expression, $S|\{n_{\!_i\gamma}\}\rangle$ is given by
(\ref{eqn:S|n>}) and $S|\{n'_{\!_i\gamma}\}\rangle$ is obtained by
replacing $n_{\!_i\gamma}$ with $n'_{\!_i\gamma}$ in
(\ref{eqn:S|n>}). Now, those orthonormal basis vectors of the form 
$|\{n_{\!_i\tau},n_{\!_i\rho},n_{\!_i\sigma}\}\rangle$ that have a
non-zero inner product with $S|\{n_{\!_i\gamma}\}\rangle$ must also be
of the form (\ref{non-zero-term}). Thus, 
$T_{\{ n_{\!_{i}\rho}\} \{ n'_{\!_{i}\rho}\} }
^{\{ n_{\!_{i}\gamma}\} \{ n'_{\!_{i}\gamma}\} }$ vanishes unless there 
exist such a set of integers $\{n_i, n'_i\}$ $(i=1,2,\cdots )$ that
\begin{eqnarray}
 n_{i}&=&n'_{i}\nonumber\\
 n_{i}+n_{\!_{i}\gamma}-n_{\!_{i}\rho}
      &=&n'_{i}+n'_{\!_{i}\gamma}-n'_{\!_{i}\rho}
\end{eqnarray}
for $\forall i$. The existence of $\{n_i\}$ and $\{n'_i\}$ is
equivalent to the condition 
$n_{\!_{i}\gamma}-n'_{\!_{i}\gamma}=n_{\!_{i}\rho}-n'_{\!_{i}\rho}$ 
for $\forall i$.
\end{proof}
\QED


\newpage

\begin{thebibliography}{99}
\bibitem{GSL}
J. D. Bekenstein, Phys. Rev. {\bf D7}, 2333(1973)
\bibitem{Frolov&Page}
V. P. Frolov, D. N. Page, Phys. Rev. Lett. {\bf 71}, 3902 (1993) 
\bibitem{Sorkin}
R. D. Sorkin, Phys. Rev. Lett. {\bf 56}, 1885(1986)
\bibitem{Panangaden&Wald}
P. Panangaden, R. M. Wald, Phys. Rev. {\bf D16}, 929(1977) 
\bibitem{Wald}
R. M. Wald, Commun. math. Phys. {\bf 45}, 9(1975); Phys. Rev. {\bf
D13}, 3176(1976)
\bibitem{1st-law}
R. M. Wald, Phys. Rev. {\bf D48}, R3427(1993); V. Iyer, R. M. Wald,
Phys. Rev. {\bf D50}, 846(1994)

\end{thebibliography}
\end{document}